\documentclass[aps,preprint]{revtex4}
\usepackage{graphicx}
\begin{document}

\title{Generation of squeezed states of microwave radiation in a superconducting resonant  circuit}

\author{A.M. Zagoskin$^{1,2,3}$}
\author{E. Il'ichev$^4$}
\author{M. W. McCutcheon$^5$}
\author{Jeff Young$^2$}
\author{Franco Nori$^{3,6}$}
\affiliation{%
(1) Department of Physics, Loughborough University, Loughborough, Leics LE11 3TU, UK}
\affiliation{(2) Department of Physics and Astronomy, The University of British Columbia, Vancouver, BC, V6T 1Z1, Canada }
\affiliation{(3) Digital Materials Laboratory, Frontier Research System, RIKEN, Wako-shi, Saitama 351-0198, Japan}
\affiliation{(4)
Institute of Photonic Technology, P.O. Box 100239, D-07702 Jena,
Germany}
\affiliation{(5) School of Engineering and Applied Sciences,  Harvard University, 
Cambridge, MA 02138,  USA}
\affiliation{(6) Center for Theoretical Physics, Physics Department, Center for the Study of Complex Systems, The University of Michigan, Ann Arbor, MI 48109-1040, USA}

\begin{abstract}
High-quality superconducting oscillators  have been successfully used for quantum control and readout devices in conjunction with superconducting qubits. Also, it is well known that squeezed states can  improve the accuracy of measurements to subquantum, or at least subthermal, levels. Here we show theoretically how to produce squeezed states of microwave radiation  in a
superconducting oscillator with tunable parameters. The circuit
impedance, and thus the resonance frequency, can be changed
by controlling the state of an RF SQUID inductively coupled to the oscillator. By repeatedly shifting the resonance frequency between any
two values, it is possible to produce squeezed and subthermal
states of the electromagnetic field in the (0.1--10) GHz range, even
when the relative frequency change is  small. We propose   experimental
protocols for the verification of squeezed state generation, and for their use to improve the readout fidelity when such oscillators serve  as   quantum transducers.
\end{abstract}

\maketitle

The problem of quantum measurements has recently attracted   renewed
attention. In quantum mechanics,
the extraction of information from a quantum system produces an
unavoidable disturbance on it. If the object is initially in an
eigenstate of the measured observable,  a quantum non-demolition
(QND) measurement can be realized, where this disturbance is
minimal~\cite{Khalili}.

One known type of detector  for   QND measurements is the so-called parametric transducer~\cite{Khalili}. A key element of a parametric transducer is an optical or radio-frequency auto-oscillator. A transducer coupled to a quantum system of interest
is  designed such that the behaviour of the quantum system   changes
the phase and/or amplitude of its oscillations. The phase
(amplitude) shift provides information about the quantum system's dynamics. With the recent development of superconducting  qubits~\cite{NoriRev,ZagoskinRev}, this approach was successfully applied to their study. In particular,  transducers (high-quality superconducting tank circuits) were used to measure the state of superconducting flux qubits~\cite{QUBITS,METHOD}.

It is known that the noise of detectors can be decreased, even below the standard quantum noise level, by employing squeezed states~\cite{UFN,Orszag,Gardiner}. In this paper we show that a superconducting
parametric transducer allows a natural application of this approach, since it can be used both to produce squeezed states and to use them in order to minimize quantum fluctuations. An immediate result of this method's application would be a way to suppress the effective noise temperature of the next-stage amplifier, at least to the nominal temperature of the cooling chamber. We emphasize that  existing experimental techniques should be sufficient for the realization of our proposal.

Squeezed states \cite{UFN,Orszag,Gardiner} are   quantum states  in which the dispersion of
one variable can be below the standard quantum limit (SQL).
More specifically, if a system is described by a pair of
dimensionless conjugate variables, $Q$ and $P$, it is in a squeezed
state if for some times

\begin{equation}
\langle \Delta Q^2 \rangle \equiv \langle Q^2 \rangle - \langle Q \rangle^2 < 1/2.
\end{equation}

The  uncertainty principle  requires that $ \langle \Delta P^2
\rangle\langle\Delta Q^2 \rangle \geq 1/4$, so when one variable is squeezed, the dispersion of
the conjugate variable increases. Squeezed states were introduced
in the context of quantum optics (see, e.g., Refs.~[\onlinecite{UFN,Orszag,Gardiner}]), but were since investigated in a
number of other systems, including polaritons ~\cite{Artoni,Hu96},
phonons ~\cite{HuNoriPhysA,Hu_phonons}, Josephson junctions \cite{HuNoriThesis} and molecular oscillations
~\cite{Janszky90}. Squeezed states have  been successfully   generated in a Josephson parametric amplifier~\cite{Yurke89}. A special interest in these states is due to
their usefulness in obtaining SQL resolution in imaging and
measurement (see, e.g., ~\cite{UFN,Orszag}). Their
classical analog can be used to obtain subthermal resolution in
mechanical measurements ~\cite{Rugar91}.

As described in, e.g., Ref.~\cite{Graham},  if a harmonic oscillator
is in a coherent state (i.e., a state with equal and minimal
uncertainties:  $ \langle \Delta P^2 \rangle = \langle\Delta Q^2
\rangle = 1/2$), a sudden change of the oscillator frequency would
create a squeezed state (while an adiabatic change would not). The
degree of squeezing is given by the ratio \begin{equation} \lambda
= \omega/\omega_{0}
\end{equation}
of the oscillator frequencies before and after the shift, or its
inverse, whichever is less than one. This topic was further
investigated in some detail in
Refs.~[\onlinecite{Agarwal91,Janszky92,Abdalla93,Kiss94,Averbukh94}].
In particular, it was shown~\cite{Janszky92} that, by repeated and properly timed
oscillator frequency shifts, one could reach an arbitrary degree
of squeezing, even for $\lambda$ close to unity (neglecting
damping, and assuming the frequency shifts to be
instantaneous). A general analysis of this
situation, under less restrictive assumptions, was given in
Ref.~[\onlinecite{Averbukh94}].

In this paper we consider the possibility of using repeated
frequency shifts to produce GHz squeezed states in a
superconducting resonant tank circuit, and the use of these states to improve the sensitivity of such a circuit when employed as a parametric transducer.
The parameters of the circuit can be tuned by controlling the state of an RF SQUID
inductively coupled to the superconducting oscillator (resonant tank circuit)~\cite{QUBITS,METHOD}.  This distinguishes our proposal from Ref.~\cite{everitt}, where the generation of squeezed states in an RF SQUID with a tunable junction was proposed through a single-step change of the parameters of the junction.

The assumption of an instantaneous  switching of the oscillator
frequency is convenient for a proof-of-principle analysis, but not
sufficient for the discussion of an experimental realization of
the effect. Therefore, we start from the density matrix $\rho$ of the
tank circuit coupled to the superconducting oscillator,   which satisfies the equation
\begin{equation}
i\partial_t \rho = [H(t),\rho],
\end{equation}
where $H(t)$ is the Hamiltonian. Additional Lindblad terms in the
r.h.s. can be added to account for dephasing and relaxation
~\cite{Gardiner}.

Initially, $H(0) = (\omega_0/2)(a_0^{\dag} a_0+a_0a_0^{\dag})$. A
change of the oscillator frequency, $\omega_0 \to \omega$,
transforms the creation/annihilation operators to ~\cite{Graham}

\begin{eqnarray*}
a = \frac{\omega+\omega_0}{2\sqrt{\omega\omega_0}}a_0 +
\frac{\omega-\omega_0}{2\sqrt{\omega\omega_0}}a_0^{\dag};\:\:
a^{\dag} =
\frac{\omega+\omega_0}{2\sqrt{\omega\omega_0}}a_0^{\dag} +
\frac{\omega-\omega_0}{2\sqrt{\omega\omega_0}}a_0.
\end{eqnarray*}

 These formulas follow from the requirement that the momentum and
position operators do not change.  This Bogoliubov transformation
can be rewritten~\cite{Graham} as 
$
a_0 \to a=Ua_0U^{\dag},
$
where
\begin{equation}
U=\exp[(1/4)\ln(\omega/\omega_0)(a_0^{\dag\:2} - a_0^2)]. \label{eq_U}
\end{equation}

The transformation of the Hamiltonian under the unitary
transformation (\ref{eq_U}) is given by
\begin{equation}
H=UH_0U^{\dag}-iU\partial_tU^{\dag};
\end{equation}
here $H_0 \equiv H(0)$ is the initial Hamiltonian; the last term is the most important in the case of fast frequency
changes (see, e.g., ~\cite{Ivanov_paper}). The resulting time-dependent
Hamiltonian, expressed in terms of the original
creation/annihilation operators, is
\begin{equation}
H(t) = H_0 +
\frac{\omega(t)^2-\omega_0^2}{4\omega(t)}(a_0^{\dag}a_0 +
a_0a_0^{\dag} + a_0^{\dag\:2} + a_0^2) - i
\frac{\dot{\omega}(t)}{\omega(t)}(a_0^{\dag\:2} - a_0^2).
\end{equation}

Finally, by moving to the interaction representation with respect
to $H_0$, we find the Hamiltonian (which, for the sake of
briefness, is also denoted by $H(t)$)
\begin{equation}
H(t) = \frac{\omega(t)^2-\omega_0^2}{4\omega(t)}(a_0^{\dag}a_0 +
a_0a_0^{\dag}) +
\frac{\omega(t)^2-\omega_0^2}{4\omega(t)}(a_0^{\dag\:2}e^{2i\omega_0t}
+ a_0^2e^{-2i\omega_0t}) - i
\frac{\dot{\omega}(t)}{\omega(t)}(a_0^{\dag\:2}e^{2i\omega_0t} -
a_0^2e^{-2i\omega_0t}).
\end{equation}

Hereafter, it is convenient to use the coherent state representation ~\cite{Orszag,Gardiner}. A coherent state $|\alpha\rangle$ is the eigenvector of the annihilation  operator with the (complex) eigenvalue $\alpha$: $|\alpha\rangle$:  $a_0|\alpha\rangle = \alpha|\alpha\rangle,$  while  $\langle\alpha|a_0^{\dag} = \langle\alpha|\alpha^*.$   Each coherent state  is a superposition of an infinite number of Fock states (states with a definite number of photons) and in the classical limit becomes a classical state with definite, time-dependent position and momentum (or other appropriate pair of canonically-conjugate classical variables).

A density matrix $\rho$ can be
represented by the Wigner function $W(\alpha,\alpha^*)$.
  Examples are a coherent state:
$
\rho_{\zeta} = |\zeta\rangle\langle\zeta| \:\leftrightarrow\:
W_{\zeta}(\alpha,\alpha^*) = \frac{2}{\pi}e^{-2|\alpha-\zeta|^2};
$
a thermal state:
$
W_T(\alpha,\alpha^*) =
\frac{2}{\pi}\tanh\left(\frac{\omega_0}{2T}\right)\exp\left[-2|\alpha|^2\tanh\left(\frac{\omega_0}{2T}\right)\right];
$
and a number state:
$
\rho_n = |n\rangle\langle n| \:\leftrightarrow\:
W_{n}(\alpha,\alpha^*) = \frac{2(-1)^n}{\pi}e^{-2|\alpha|^2} {\rm
L}_n(4|\alpha|^2).
$
Here ${\rm L}_n(z)$ is the Laguerre polynomial.  The complex variables $\alpha$ and $\alpha^*$ can be expressed through their real quadrature components,
$x,y$:  $\alpha = x + iy$, $\alpha^* = x - iy$, which can be related to directly observable properties of an oscillator (e.g., current and voltage). For our purposes, the usefulness of this representation is due to
the fact that there exists a one-to-one correspondence~\cite{Gardiner}   between the action of
creation/annihilation operators on $\rho$ and differential
operations on $W(\alpha,\alpha^*)$:
\begin{eqnarray*}
a_0\rho \leftrightarrow [\alpha +
(1/2)\partial_{\alpha^*}]\:W(\alpha,\alpha^*);\: \rho a_0
\leftrightarrow [\alpha -
(1/2)\partial_{\alpha^*}]\:W(\alpha,\alpha^*);\\
 a_0^{\dag}\rho \leftrightarrow [\alpha^* - (1/2)\partial_{\alpha}]\:W(\alpha,\alpha^*);\:
 \rho a_0^{\dag} \leftrightarrow [\alpha^* + (1/2)\partial_{\alpha}]\:W(\alpha,\alpha^*).
 \end{eqnarray*}
Therefore the Liouville equation for the density matrix operator
is now replaced by a partial differential equation for a c-number Wigner function
 \begin{eqnarray}
i\partial_tW(\alpha,\alpha^*) = 2
\beta(t)\:[\alpha^*\partial_{\alpha^*}-\alpha\partial_{\alpha}]\:
W(\alpha,\alpha^*) +
2\:[\gamma(t)^*\alpha\partial_{\alpha^*}-\gamma(t)\alpha^*\partial_{\alpha}]\:W(\alpha,\alpha^*) \label{eq_W}
 \end{eqnarray}
 Here we have introduced
\begin{eqnarray}
\beta(t) = \frac{\omega_0}{4} \left[\left(\frac{\omega(t)}{\omega_0}\right)^2-1\right]
\equiv \frac{\omega_0}{4} \left[\lambda(t)^2-1\right];\\
\gamma(t) = \left[\beta(t)+i\frac{\dot{\omega}(t)}{\omega(t)}\right]e^{2i\omega_0t}
 \equiv \left[\beta(t)+i\frac{\dot{\lambda}(t)}{\lambda(t)}\right]e^{2i\omega_0t}.
\end{eqnarray}

In the presence of dissipation, Eq.~(\ref{eq_W}) will also contain
second-order terms, and would be only tractable numerically.
Neglecting these terms leads to an immediate simplification, since
Eq.~(\ref{eq_W})   is a differential equation of
first order and can be solved by the method of characteristics. The characteristics $x(t), y(t)$ satisfy the equations:
\begin{eqnarray}
\frac{1}{2}\dot{x} = [{\rm Im}\gamma(t)] \:x(t) + \left\{\beta(t)-[{\rm
Re}\gamma(t)]\right\} \:y(t); \nonumber\\
\\\label{eq_char}
\frac{1}{2}\dot{y} =   -\left\{\beta(t)+[{\rm Re}\gamma(t)]\right\} \:x(t)  -
[{\rm Im}\gamma(t)] \:y(t). \nonumber
\end{eqnarray}
After finding the solutions $x(x_0,y_0,t); y(x_0,y_0,t)$ (where
$x_0, y_0$ are the initial conditions) and inverting them to
obtain $x_0(x,y,t); y_0(x,y,t)$, one obtains the Wigner function
for an arbitrary point and time from its value at $t=0$ via
 \begin{equation}
 W(x+iy,x-iy,t) =  W(x_0(x,y,t)+iy_0(x,y,t),x_0(x,y,t)-iy_0(x,y,t),0). \label{eq_drag}
 \end{equation}
Eq.~(\ref{eq_drag}) provides the complete formal solution for the
quantum mechanical problem of a harmonic oscillator with variable
frequency in terms of the characteristics of Eq.~(\ref{eq_W}), as should be
expected~\cite{UFN,exact solution}. Equations (\ref{eq_char})   are
of the classical Hamilton type, and can be  solved in general only
numerically. Still, a good analytical approximation can be found under two assumptions. First, the relative change of the oscillator frequency must be
small ($|1-\lambda| \ll 1$). (This assumption holds for any
reasonable experimental realization.) Second, the frequency must change either
very fast ($\dot{\lambda} \gg \omega_0$), or very slowly, ($\dot{\lambda} \ll \omega_0$), compared to the oscillator period. It was shown~\cite{Averbukh94} in a general case, that only when the frequency is changed fast in one direction and slowly in the other, an arbitrarily strong squeezing by small repeated frequency shifts can take place. Therefore our analytical approximation is good for describing the very regime we are interested in.

 Let us first treat the fast limit. In this case, $\beta(t)$ can be
neglected compared to $\gamma(t)$, and the   equations
(\ref{eq_char}) are reduced to
$
\dot{\alpha} = -2i\gamma(t)\alpha^*(t);$ 
$\:\:
\dot{\alpha}^* = 2i\gamma(t)^*\alpha(t).
$
We must keep  all terms in $\gamma(t)$ until we transform this
system into a second-order equation for $\alpha(t)$. Now, dropping
the small terms (assuming $\ddot{\lambda} \ll \dot{\lambda} \equiv
v$), we obtain

\begin{equation}
\ddot{\alpha} + v \dot{\alpha} - 4v^2\alpha = 0.
\end{equation}

\begin{figure}[h]
\includegraphics
[width=9cm]{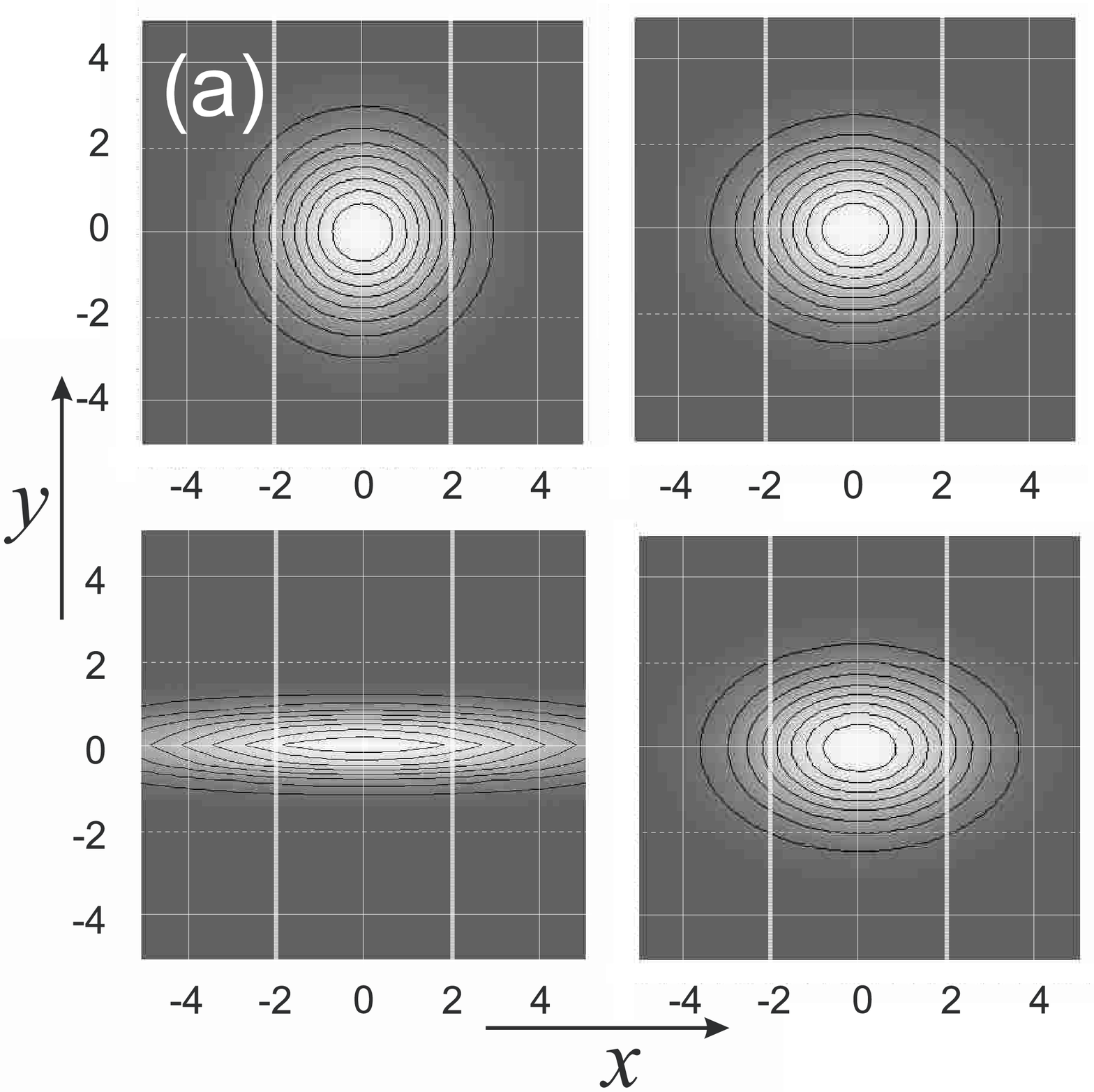}\\
\includegraphics
[width=9cm]{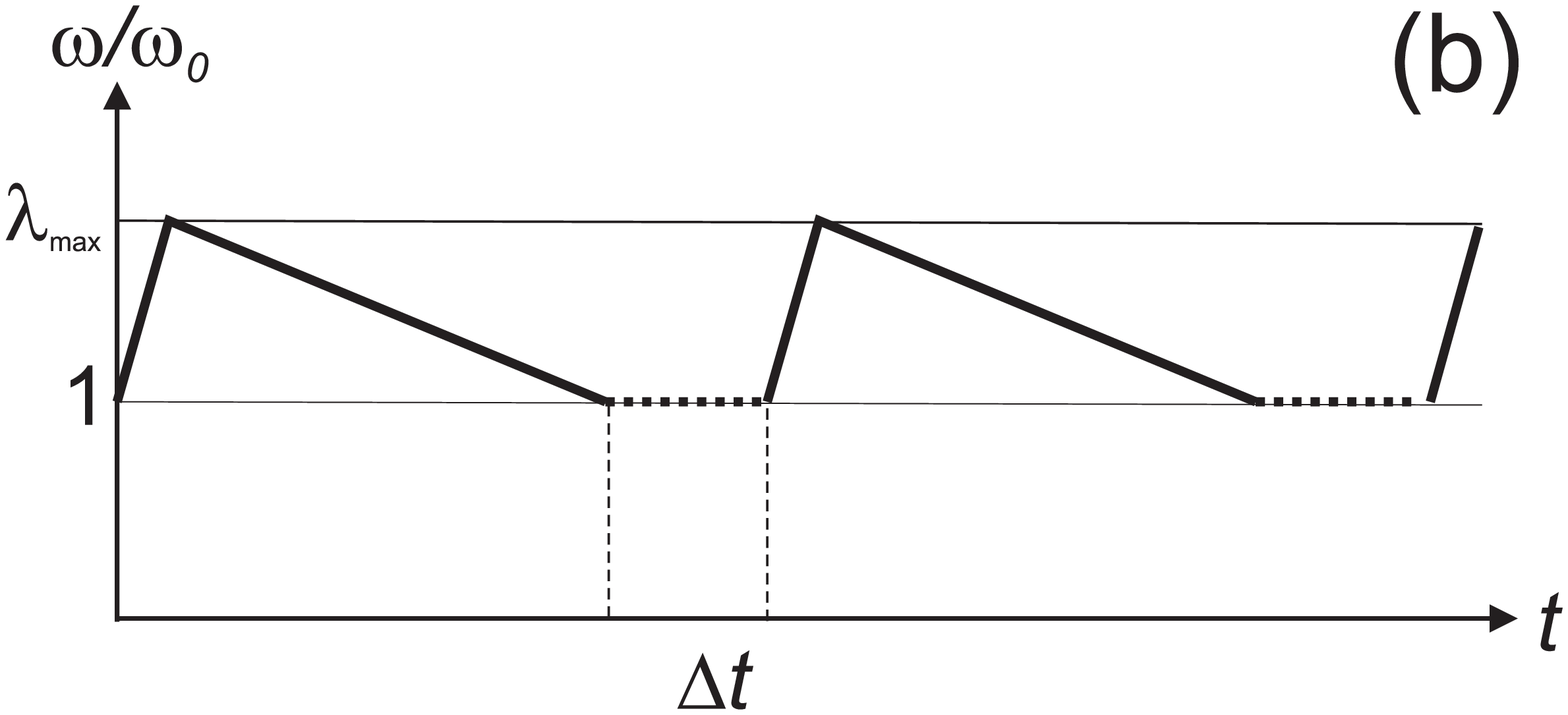}
\caption{Squeezing of a thermal state, with $k_B T = 4 \omega_0$,  by repeated frequency shifts, with $\lambda_{\rm max}-1 = 0.05$. (a) Contour plots of the approximate expression of the Wigner function in the interaction representation (see Eqs.~(12,15,17)) are shown as a function of the quadrature variables, $x$ and $y$. Clockwise starting from the upper left: initial thermal state, the state after one, two, and ten cycles. The dark background corresponds to $W=0$, and white to $W=0.08$.
(b) A schematic frequency-versus-time dependence necessary to produce the results of Fig.~1a. The idle periods $\Delta t$ are chosen to ensure that every fast shift occurs in the same phase with respect to the quadrature coordinates. Here one must  account  for the fact that in the Schroedinger representation the whole phase plane rotates around the origin with the base oscillator frequency $\omega_0$. An additional rotation of the Wigner function as a whole during the slow shift, Eq.~(\ref{eq_slow_rot}), can be neglected in comparison, since the additional phase $\left|\delta\theta\right| \leq 0.5~\omega_0~t \left|\lambda_{\rm max}^2-1\right| \ll  ~\omega_0~t.$
}\label{fig1}
\end{figure}

In the case of a linear frequency change, $v =$ const, this equation is
easily solved with the initial conditions $\alpha(0) = \alpha, \:
\dot{\alpha}(0) = 2v\alpha^*$, yielding (see Fig.~\ref{fig1})

\begin{eqnarray}
x_0(t,x,y) = x_0(t,x) = \frac{x~\sqrt{17}}{(2-s_-/v)e^{s_+t}+(s_+/v-2)e^{s_-t}};\\
y_0(t,x,y) = y_0(t,y) = \frac{y~\sqrt{17}}{(s_+/v-1)e^{s_+t}+(1-s_-/v)e^{s_-t}};\\
s_+ = v\left(\sqrt{17}-1\right)/2; \:\:\:\:\:\: s_- = -v\left(\sqrt{17}+1\right)/2.
\end{eqnarray}

In the slow regime, the terms with $\dot{\lambda}$ can be
neglected. The remaining terms are of the same order, but
$\gamma$-terms are  very fast (oscillating with
$2\omega_0$) and average to zero over the period of the oscillator.
Therefore we are left with
\begin{eqnarray}
\frac{1}{2}\dot{x} = \beta(t) y(t); \:\:\:\:\:\:
 \frac{1}{2}\dot{y} =   -\beta(t) x(t), \label{eq_char_slow}
 \end{eqnarray}
  or \begin{equation}
\dot{\alpha} = -2i\beta(t) \alpha(t), \Longrightarrow \alpha(t) =
\alpha(0) \exp\left[-i\frac{\omega_0}{2} \int_0^t (\lambda^2(s)-1) ds\right]. \label{eq_slow_rot}
\end{equation}
Even without solving these equations, it is clear that the slow
regime cannot affect squeezing in any way: Eq.~(\ref{eq_char_slow})
describes circles in   phase plane; the evolution of the Wigner
function given by (\ref{eq_drag}) therefore reduces to its (slow
compared to $\omega_0$) rotation as a whole, without changing
shape. This conclusion is consistent with previous work~\cite{Graham,Averbukh94}.

The experimental realization of this proposal is rather challenging.
There have been several reports of an ultrafast 
perturbation of optical microcavity modes using the 
dispersion of injected free carriers~\cite{Fushman, Preble, McCutcheon}.
However, the small frequency shifts (of the order $|\lambda -1| \sim 5 \times 10^{-4}$) were on a picosecond timescale, which is slow with respect to the period of one optical cycle of the cavity modes.   These processes are therefore in the adiabatic limit, 
rather than in the sudden-frequency-shift regime~\cite{Graham}. Thus, while they may prove useful for on-chip frequency
conversion~\cite{Notomi06, NotomiPRA}, they are
not useful for generating non-classical optical states. Moreover, there is little prospect for a repeated application of the perturbation within the sub-nanosecond lifetime of a microcavity mode.

In optical lattices occupied by cold atoms, 
the relatively low oscillation frequencies ($\sim$ 1 MHz) and
precise dynamic control of the atomic potentials have allowed the 
demonstration of squeezed positional states~\cite{Rudy}.
The situation is also promising in the (0.1--10)GHz range, where one can 
use  Josephson
junctions as nondissipative nonlinear elements, allowing 
control of the circuit parameters.

If the frequency is low, so that $\hbar\omega\simeq$ $k_B T$, where
$T$ is 10--50 mK (dilution refrigerator temperatures) the
so-called RF SQUID~\cite{bar} configuration can be used. The system
consists of a high-quality superconducting tank circuit,
inductively coupled to a loop containing the Josephson contact
(the single junction interferometer~\cite{bar}). In the dispersive
mode of an RF SQUID, the effective inductance of the system
becomes~\cite{bar}:
\begin{equation}
L_{\rm eff} = L_T - \frac{M^2}{L + {\cal L}(\varphi)},
\end{equation}
where $L_T$, $L$ are respectively the self-inductance of the tank
circuit and of the RF SQUID loop, $M$ is their mutual inductance,
and
\begin{equation}
{\cal L}(\varphi) = \frac{L}{\beta \cos \varphi}
\end{equation}
is the Josephson inductance of the SQUID junction, which depends
on the phase bias $\varphi$ across it. Here $\beta \equiv 2 \pi
LI_{C}/\Phi_0 $ and $I_{C}$ is the critical current of the SQUID
Josephson junction. Therefore, by varying $\varphi$ one controls
the effective inductance and eigenfrequency of the tank:
\begin{equation}
\frac{\omega_{\rm eff}^2}{\omega_0^2} = \frac{L}{L_{\rm eff}} \approx 1 +
\frac{k^2}{(1 + \beta \cos \varphi)}.
 \end{equation}
Here $k^2 \equiv M^2/LL_T \ll 1$ (in practice $k \approx 0.3$ can
be easily realized) is the coupling coefficient between the tank and the
SQUID loop. Since for a SQUID dispersive mode $\beta < 1$, the
variations $\delta
\omega_0$ of the eigenfrequency of the tank satisfy $\delta
\omega_0\simeq (0.1-0.01) \omega_0$, which should be enough for our
purposes.

For higher frequencies $\hbar\omega \gg$ $k_B T$, a tuneable
superconducting cavity could be used. In such a device, a DC
SQUID is incorporated to the strip resonator~\cite{chal} and $\delta
\omega_0\simeq (0.1-0.01) \omega_0$ as well.

For both cases, the phase-changing pulse must be sharp on the scale
of $\omega_0$ and this is within the current experimental
capabilities. The coherence time of the system is currently in the range of 1--10 $\mu$s, allowing for at least several cycles of
frequency change, with the corresponding increase in the
squeezing of the final state.

In conclusion, we have shown that the above procedure can produce
squeezed states in an nonlinear superconducting oscillator, so
that the fluctuations of the amplitude (phase) of the oscillator
are suppressed along certain directions in phase space, which rotate
with  the base oscillator frequency $\omega_0$. By making
use of this noise suppression, the measurements of the amplitude (phase) of
these oscillators can reach a sensitivity below the standard quantum limit, or at
least, below the thermal level.

\acknowledgements AZ is grateful to B. Ivanov for valuable discussions and
for pointing out  references ~\cite{Ivanov_paper,exact
solution}. FN gratefully acknowledges partial support from the National
Security Agency (NSA), Laboratory Physical Science (LPS), Army
Research Office (ARO), National Science Foundation (NSF) grant
No. EIA-0130383, JSPS-RFBR 06-02-91200, and Core-to-Core (CTC)
program supported by the Japan Society for Promotion of Science (JSPS).

\end{document}